# Measuring the linear and non-linear elastic properties of brain tissue with shear waves and inverse analysis

Yi Jiang[1,&], Guoyang Li[1,&], Lin-Xue Qian[2], Si Liang[2], Michel Destrade[3]*, Yanping Cao[1]*

[1]Institute of Biomechanics and Medical Engineering, AML, Department of Engineering Mechanics, Tsinghua University, Beijing 100084, P. R. China

[2]Department of Ultrasound, Beijing Friendship Hospital, Capital Medical University, Beijing 100050, P. R. China

[3]School of Mathematics, Statistics and Applied Mathematics, National University of Ireland Galway, University Road, Galway, Ireland

**Abstract**: We use supersonic shear wave imaging (SSI) technique to measure not only the linear but also the non-linear elastic properties of brain matter. Here we tested six porcine brains *ex vivo* and measured the velocities of the plane shear waves induced by acoustic radiation force at different states of pre-deformation when the ultrasonic probe is pushed into the soft tissue. We relied on an inverse method based on the theory governing the propagation of small-amplitude acoustic waves in deformed solids to interpret the experimental data. We found that, depending on the subjects, the resulting initial shear modulus $\mu_0$ varies from 1.8 kPa to 3.2 kPa, the stiffening parameter *b* of the hyperelastic Demiray-Fung model from 0.13 to 0.73, and the third- (*A*) and fourth-order (*D*) constants of weakly non-linear elasticity from -1.3 kPa to -20.6 kPa and from 3.1 kPa to 8.7 kPa, respectively. Paired *t*-test performed on the experimental results of the left and right lobes of the brain shows no significant difference. These values are in line with those reported in the literature on brain tissue, indicating that the SSI method, combined to the inverse analysis, is an efficient and powerful tool for the mechanical characterization of brain tissue which is of great importance for computer simulation of traumatic brain injury and virtual neurosurgery.

*Corresponding authors:
Yanping Cao (Y. P. Cao), caoyanping@tsinghua.edu.cn; Michel Destrade, michel.destrade@nuigalway.ie
[&]These authors made equal contribution to this study.

# 1. Introduction

The determination of the mechanical properties of brain tissue is essential to the modeling, simulation and understanding of its behavior when deformed by various external and internal stimuli. For instance, computer simulation of traumatic brain injury and virtual neurosurgery requires a precise knowledge of the detailed mechanical parameters of brain tissue (Kleiven and Hardy 2002; Miga and Paulsen 2000; Miller 1999; Roberts et al. 1999; Zhang et al. 2001). Moreover, some diseases, such as hydrocephalus and Alzheimer's disease, are known to alter the mechanical properties of the brain (Kruse et al. 2008). Therefore, measuring accurately the mechanical properties of brain tissue may help the diagnosis of these diseases and monitor their development. Finally, determining the exact contours of a brain tumor is not an exact science at present, and neurosurgeons often rely on intuition and experience to evaluate them through direct palpation (Macé et al., 2011). However tumors have mechanical properties which are notably different from the surrounding healthy tissue, and being able to obtain an accurate elastic stiffness map of a brain region would be a major advance of medicine.

Numerous efforts have been made to determine the mechanical properties of brain matter. The extensive review by Chatelin et al. (2010) summarizes the main efforts devoted to this vital issue from the 1960s to 2010. The testing methods include the conventional tensile, compression and shear tests (Hrapko et al. 2006; Karimi et al. 2013; Miller and Chinzei 1997; Miller et al. 2002; Nicolle et al. 2005; Pervin and Chen 2009; Prange and Margulies 2002; Rashid et al. 2013a, 2013b; Saraf et al. 2007), which are carried out on samples cut out of the organ. In contrast, the indentation, the pipette aspiration method, and the magnetic resonance elastography (MRE) method allow us to conduct the tests on intact brains. In particular, indentation tests and pipette aspiration methods have been frequently used to determine the *linear* elastic modulus of the brain (Miller et al. 2000; Gefen et al. 2003; Gefen and Margulies 2004; Kaster et al. 2011; Prevost 2011). Some authors (Kaster et al. 2011; Miller et al. 2000) have used the indentation method to determine the *hyperelastic* properties of the brain tissue as well. However, the inverse analysis used to extract the hyperelastic (non-linear) parameters from indentation or pipette aspiration tests can be over-sensitive to data errors according to some recent studies (Zhang et al. 2014a, 2014b). Besides, it is impossible to probe the mechanical properties of the interior part of the brain using indentation or pipette aspiration tests without destructing the sample. The MRE method is very promising, as it permits the evaluation of the *in vivo* mechanical properties of the brain tissue non-invasively. Hence a number of authors have determined the linear elastic and linear viscoelastic properties of brain tissue using the MRE method (Atay et al., 2008; Klatt et al., 2007; Kruse et al., 2008; Green et al., 2008; Sack et al., 2009; Streitberger et al., 2010). However, the spatial resolution of this technique is limited due to long acquisition time, as pointed out by Macé et al. (2011).

In this study, we measured not only the linear but also the non-linear elastic properties of brain tissue using the *supersonic shear imaging* (SSI) technique (Bercoff et al., 2004a) and an inverse analysis based on the theory of the *large acousto-elastic effect* (Ogden 2007; Destrade et al. 2010b).

The SSI method has been proposed and commercialized quite recently to measure the velocity of the small-amplitude plane shear wave generated by acoustic radiation force in soft tissue. Such a wave propagates in an undeformed, isotropic, incompressible, homogeneous, elastic material with speed $c$ given by

$$\rho c^2 = \mu_0, \qquad (1)$$

where $\rho$ is the (constant) mass density and $\mu_0$ is the initial shear modulus of linear (second-order) incompressible elasticity. In that sense, measuring the wave speed is equivalent to measuring the stiffness of the solid. For instance, Bercoff et al. (2004a) used this equation to produce an "elasticity mapping" of some phantom gels and of *in vivo* human breast tissue.

However, brain tissue is not undeformed *in situ*: it is slightly compressed by its encasing inside the *dura mater*. The same holds true for most living tissues which are usually subjected to residual and initial stresses for optimal response to mechanical stresses. It is thus important to access material properties at the next order, beyond the hypothesis of linearity. The theory of *acousto-elasticity* is at hand to help us with this task. Dating back to the works of Brillouin (1946), it provides a direct access to $A$, the (Landau) third-order elasticity constant, by giving the speed of the shear wave when the solid is subject to a small uni-axial elongation $e$ ($e > 0$ in tension, $e < 0$ in compression) as (Destrade et al., 2010b)

$$\rho c^2 = \mu_0 + \frac{A}{4} e. \qquad (2)$$

This is the squared speed of the shear wave propagating normal to the direction of elongation, with polarization along that direction, which, as explained by Jiang et al. (2014), is the wave relevant to SSI imaging. For instance, Genisson et al. (2007), Rénier et al. (2008) and Larotte et al. (2012) used acousto-elasticity to determine the second- and third-order elastic constants of phantom gels and of beef liver samples.

When larger deformations of tissues are to be accounted for, such as those encountered in trauma, swelling or surgery for instance, fourth-order elasticity must be invoked. It requires knowledge of one more elastic constant, $D$ (Hamilton et al., 2004; Destrade and Ogden, 2010). Rénier et al. (2008) have evaluated this constant for some phantom gels by first relying on an acousto-elastic experiment and Eq. (2), and next by performing a *separate* experiment involving finite amplitude shear waves. Here we rely instead on the *large acousto-elastic effect*, which extends Eq. (2) to fourth-order elasticity as

$$\rho c^2 = \mu_0 + \frac{A}{4} e + (2\mu_0 + A + 3D)e^2. \qquad (3)$$

Here, all is required is to record the quadratic variation of the squared wave speed with the elongation, using a *single* experiment. As soon as it ceases to vary linearly, we enter the realm of large acousto-elasticity, and by inverse analysis of the coefficients of the quadratic we can access all three elastic constants.

Finally, it could well be that the behavior of brain matter under very large deformation is of interest to medicine, say for the study of large compressions due to hemorrhage, traumatic brain injury swelling, tumor growth, or neuro-surgery. In that case, the incremental modeling of weakly nonlinear elasticity (sequence of second-, third-, fourth-order expansions) must be abandoned and finite elasticity theory must be invoked. In that framework, the soft material is seen as hyperelastic, with a strain energy density $W$ being a symmetric function of the principal stretches: $W = W(\lambda_1, \lambda_2, \lambda_3)$. Here the theory of *exact acousto-elasticity* provides a link between the shear wave speed and *W* through the following formula (Ogden, 2007)

$$\rho c^2 = \frac{\lambda_2^2 \left( \lambda_1 \frac{\partial W}{\lambda_1} - \lambda_2 \frac{\partial W}{\partial \lambda_2} \right)}{\lambda_1^2 - \lambda_2^2},\tag{4}$$

although now the resulting inverse analysis is nonlinear in general. Here, $\lambda = 1 + e$ is the principal stretch along the direction of elongation (and *e* can be finite), while $\lambda_2, \lambda_3$ are the lateral principal stretches.

For this paper we performed *ex vivo* experiments on five porcine brains using the Supersonic Imagine device, a commercial implementation of the SSI technique. Section 2 presents the detailed procedure together with a brief overview of the SSI technique. In Section 3 we analyze the experimental data and determine the linear and non-linear elastic constants of porcine brain matter using an inverse analysis. In particular for the modeling relying on weakly nonlinear elasticity theory, we modify Eq. (3) to account for a homogeneous pre-deformation which would be not exactly uni-axial (Jiang et al., 2014). For the modeling of brain matter relying on finite elasticity, we specialize Eq. (4) to the widely used Demiray-Fung strain energy density. Finally, we highlight and discuss the merits and limitations of the present method in the final two sections.

The resulting linear and non-linear elastic coefficients are found to be in line with those reported in the literature on brain matter (Rashid et al. 2013b) and to be physically sound. They confirm that brain matter is a *very soft material*, even softer than a tissue mimicking gelatin (Rénier et al. 2008).

## 2. Experimental Methods

Experiments were conducted on the porcine brain tissue to measure its linear and non-linear elastic properties. The Aixplorer® ultrasound instrument (Supersonic Imagine, Aix-en-Provence, France) was used, upon which the SSI technique (Bercoff et al. 2004a) has been implemented. A brief introduction on the SSI technique and the details of the materials and measurements are given below.

### 2.1 SSI technique

The Supersonic shear imaging (SSI) technique uses the focused ultrasonic beam to create the acoustic radiation force, which is due to the momentum transfer from the acoustic wave to the medium, and successively focuses the ultrasonic beam at different depth (Bercoff et al. 2004a). The successively focusing beam acts as a shear source and moves at a supersonic speed in the medium, thus the resulting displacement field is confined in a Mach cone. In this case, the generated two quasi-plane shear wave fronts interfere along the Mach cone and propagate in opposite directions; this phenomenon is the elastic Cerenkov effect (Bercoff et al. 2004b). The interference between shear waves leads to a cumulative effect, which induces greater displacements in the soft tissue. Videos S1 and S2 in the Supplementary Information material and Fig. 1 illustrate the elastic Cerenkov effect (Bercoff et al. 2004b) corresponding to the following displacement field caused by a moving point force

$$u_i(\mathbf{r},t) = \frac{1}{4\pi\rho c_s^2 R\sqrt{1-M_e^2\sin^2\Theta}} \sum_{k=1}^{2}(\delta_{ij} - \frac{\partial R(\tau_k^*)}{\partial x_i}\frac{\partial R(\tau_k^*)}{\partial x_j}), \tag{5}$$

where

$$\begin{cases} \tau_1^* = t + \dfrac{R}{c_s(M_e^2-1)}(M_e\cos\Theta + \sqrt{1-M_e^2\sin^2\Theta}) \\ \tau_2^* = t + \dfrac{R}{c_s(M_e^2-1)}(M_e\cos\Theta - \sqrt{1-M_e^2\sin^2\Theta}) \end{cases} \tag{6}$$

$$R = |\mathbf{r}-\mathbf{r}_0(t)|, \mathbf{r}_0(t) = c_0 t \mathbf{e}_j, \Theta(t) = \arccos(\frac{x_j - c_0 t}{|\mathbf{r}-\mathbf{r}_0(t)|}) \tag{7}$$

Here $M_e$ is the Mach number defined as the ratio of speed of the moving shear source $c_0$ to the velocity of the generated shear wave $c_s$, $t$ is the time, $\rho$ the mass density, and $\delta_{ij}$ the Kronecker delta. Also, $x_i$ is the coordinate, $\mathbf{e}_j$ is the base vector of direction $x_j$ and in Eq. (5) the shear source moves along $x_j$ direction. In SSI technique, an ultrafast scanner prototype is adopted to image the propagation of the shear waves in less than 20ms. This technique received FDA approval in 2009 and has since been commercialized.

### 2.2 Materials

Porcine brain matter, as shown in Fig. 2, was obtained from freshly slaughtered animals. The fresh porcine brains were kept between 2°C to 4°C in an ice box and transported to the Beijing Friendship Hospital within 12 hours postmortem, where the measurements were carried out. A total of six intact porcine brains were tested. According to Rashid et al. (2013a), keeping freshly slaughtered porcine brains at ice-cold temperature prior to testing helps minimize the difference between the measured *in vitro* test results and the *in vivo* properties.

### 2.3 Measurements

Measurements were performed at room temperature. The measurements were conducted in water so that it was not necessary to apply ultrasound gel on the probe. They were conducted within minutes, to ensure that there was no tissue degradation due to the effect of water on cells. The Aixplorer® transducer SL15-4 was adopted in our experiments, used with a 9MHz central frequency.

The experimental protocol went as follows.

**(1)** Without applying any pressure, we placed the probe on the sample and measured the shear wave velocity in a region of interest (circle region in Fig. 3) using the SSI technique; **(2)** Then we imposed compression by pushing the probe into the surface of the brain and measured the shear wave speed in the deformed solid, which now exhibits strain-induced anisotropy. We recorded the stretches and the corresponding deformation of the soft tissue based on the B-mode image provided by the Aixplorer® system; **(3)** Increasing the amount of deformation by moving the probe down, we repeated the procedure of Step **(2)** to measure the corresponding shear wave speeds at different amounts of compression.

Fig. 3 shows the wave speeds recorded by the Supersonic Imagine at two different amount of compression for the left lobe of Pig #3. It can be clearly seen from the figure that the wave speed increases with the increase in the amount of compression. This information will be used to deduce the non-linear parameters of the brain tissue, as described in detail in the next section. It should be pointed out that both the shear wave velocity and the deformation of the soft tissue at different amounts of compression are measured using the Supersonic Imagine system; there is no need to measure the compression force, because we will use speed-deformation relationships such as those given by Eqs. (1)-(4), not speed-stress relationships.

The deformation of the soft tissue in the region of interest (disc in Fig. 3) is assumed to be homogeneous and quantified by the measured principal stretches along the loading and the lateral directions (details of the measurement technique for the principal stretches are given in (Jiang et al. , 2014)). The measurements were conducted on the right and left lobes separately instead of the whole brain and two characteristic points at the upper and bottom surfaces were used to evaluate the principal stretches along loading direction. The size and location of the measurement area are shown in Fig. 3: it was located in the left parietal lobe and the centre of the region of interest (a disc of radius 4 mm). Fig. 4 displays the averaged shear wave speeds with corresponding standard deviations, over the full range of compression achieved (up to almost 40%).

**3. Inverse analysis to determine linear and non-linear elastic properties**

With the velocities of the shear wave given by the Supersonic Imagine device (Fig. 4) at different amounts of compression, we can determine the linear and non-linear elastic parameters *via* an inverse analysis. To this end, we invoke the theory of small-amplitude shear wave propagation in a deformed soft material, which is covered by the theory of acousto-elasticity (Ogden 2007). The soft tissue is assumed to be incompressible, isotropic, homogeneous and non-linearly elastic (Rashid et al. 2013b). For the constitutive modeling we consider in turn two of the most used strain energy densities of non-linear elasticity.

In *Finite Elasticity, W* is a given explicit function of the strain. In that context we use the popular Demiray-Fung model (Demiray, 1972), with strain energy function

$$W = \frac{\mu_0}{2b}(e^{b(I_1-3)} - 1), \qquad (8)$$

where $\mu_0$ and $b$ are positive material parameters: $\mu_0$ represents the initial shear modulus and $b$ defines the strain-stiffening behavior. Also, $I_1 = \lambda_1^2 + \lambda_2^2 + \lambda_3^2$, where $\lambda_1$ is the principal stretch along the loading direction and $\lambda_2$, $\lambda_3$ are the principal stretches along the lateral directions. The incompressibility condition requires that no volume change takes place, so that $\lambda_1 \lambda_2 \lambda_3 = 1$. In the region of interest, we let $\lambda_1 = \lambda$ and $\lambda_2 = \lambda_1^{-\xi}$, $\lambda_3 = \lambda^{-(1-\xi)}$, where $\xi$ is a tri-axiality parameter in the range of 0-1. Based on the theory of small-amplitude shear waves propagating in deformed hyperelastic solids (Ogden 2007), Jiang et al. (2014) have derived an analytical solution for the material model given by Eq. (8) to predict the velocity $c$ of a shear wave propagating orthogonal to the direction of the loading direction, and polarized along the loading direction, as

$$\rho c^2 = \mu_0 \lambda^{-2\xi} e^{b\left(\lambda^2 + \lambda^{-2\xi} + \lambda^{-2(1-\xi)} - 3\right)} \qquad (9)$$

where $\rho$ is the (constant) mass density of the homogeneous incompressible soft tissue (here we took $\rho$=1.04 g/ml for brain tissue). This equation follows from the substitution of the particular choice of $W$ given by Eq.(8) into the general formula Eq.(4). We will use Eq. (2) to conduct an inverse analysis to determine the linear elastic parameter $\mu_0$ and the non-linear stiffening elastic parameter $b$ of brain tissue. The optimization process following from Eq.(9) can be rendered linear by taking the log on both sides, and it then yields a unique set of best-fit parameters, see (Jiang et al., 2014) for details. The parameter $\xi$ in the present experiments was taken as 0.5 according to the deformation state and hence two characteristic points were sufficient to evaluate the principal stretches.

In *Physical Acoustics,* the strain energy density is often expanded in terms of powers of the Green strain tensor $\mathbf{E}$, for instance as

$$W = \mu_0 tr(\mathbf{E}^2) + \tfrac{1}{3} A tr(\mathbf{E}^3) + D(tr(\mathbf{E}^2))^2, \qquad (10)$$

for fourth-order incompressible isotropic elasticity, and higher order terms are neglected (Hamilton et al. 2004, Destrade and Ogden 2010). Here $\mu_0$ is the initial shear modulus of linear elasticity, and $A, D$ are the Landau elastic constants of third- and fourth-order elasticity, respectively. In that context we write $\lambda = 1 + e$, where $e$ is the elongation, and expand the squared velocity of Eq.(4) up to second order, as

$$\rho c^2 = \mu_0 + \gamma_1 e + \gamma_2 e^2, \qquad (11)$$

where we find that

$$\gamma_1 = \tfrac{1}{2} A + 2\mu_0 - (4\mu_0 + \tfrac{1}{2} A)\xi,$$
$$\gamma_2 = \mu_0 + \tfrac{5}{4} A + 4D - (2\mu_0 + \tfrac{7}{4} A - 4D)\xi + (8\mu_0 + \tfrac{5}{2} A + 4D)\xi^2, \qquad (12)$$

in agreement with the formulas established in the case of equi-biaxial lateral expansion $\xi = 1/2$, which reduce to $\gamma_1 = A/4$, $\gamma_2 = 2\mu_0 + A + 3D$, see Eq.(3) and (Destrade et al., 2010b).

The wave velocities reported in Fig. 4 at different stages of compression indicate a non-linear relationship between the squared speed and the stretch/elongation. With the Demiray-Fung model given by Eq. (8), the initial shear modulus $\mu_0$ and the hyperelastic parameter *b* of brain tissue are determined and shown in Fig. 4. The dimensionless parameter *b* varies from 0.13 to 0.73 for different subjects. These values are comparable with those given by other independent measurements. Hence, Rashid et al. (2013b) measured the linear and non-linear elastic properties of brain tissue using simple shear tests. They obtained an initial shear modulus of 4.94 kPa and stiffening parameter *b* varying from 0.12 to 0.52.

Similarly, from fitting the data to the fourth-order elasticity model Eq.(10) and a quadratic dependence of the squared wave speed to the elongation, we get access to $\mu_0$, $\gamma_1$, $\gamma_2$ and consequently (for the first time) to values of the third- and fourth-order elastic constants *A* and *D* for brain matter, see Table 1 for the results. In particular we can compare them to those obtained by Rénier et al. (2008) for soft agar-gelatin gels. Those authors used two different protocols: first, a small acousto-elastic effect experiment to measure $\mu_0$ and *A* and second, a non-linear wave propagation to access *D*, in contrast to our single large acousto-elastic experiment. For their softest gel (5% gelatin) they found $\mu_0$ = 6.6 kPa, *A* = -37.7 kPa, *D* = 30 kPa. For a quick comparison we take the average right lobe values of Table 2: $\mu_0$ = 2.3 kPa, *A* = -13.4 kPa, *D* = 5.1 kPa, so that we can draw and compare the curves of predicted mechanical behavior for gel and brain matter when deformed. For uni-axial tension, simple shear, and pure bending, we have

$$T = 3\mu_0 e + 3(\tfrac{1}{4}A + \mu_0)e^2 + (7\mu_0 + \tfrac{15}{4}A + 9D)e^3,$$

$$S = \mu_0 K + (\mu_0 + \tfrac{1}{2}A + D)K^3 ,$$

$$M = \frac{4}{3}\kappa + \frac{2}{5}\left(1 + \frac{A + 2D}{\mu_0}\right)\kappa^3, \tag{13}$$

respectively, where *T* is the tensile Cauchy stress component and *e* the elongation; *S* is the shear Cauchy stress component and *K* the amount of shear; and *M* is a non-dimensional measure of the bending moment and $\kappa$ is the product of the bent block's aspect ratio by the bending angle (Destrade et al. 2010a). To save space we do not reproduce the resulting curves here but we note that they all show an almost linear variation of *T*, *S*, *M* over the range 0.0-0.3 for *e*, *K*, $\kappa$, respectively, with the slope of the gel curves being at least three times that of the brain curves. In that sense, we establish that porcine brain matter is three times softer than a very soft agar-gelatin gel.

As we performed measurements on both the left and right lobes of the brains, it is interesting to compare the results and find whether significant differences between the mechanical properties of the left and right lobes of the brain exist. Paired *t*-test was invoked for the statistical analysis and $p \leq 0.05$ was adopted as a criterion of significant difference.

The results are listed in Table 2 and all the values for $p$ are greater than 0.5, indicating that no significant difference exists between the mechanical properties of the left and right lobes of the brain.

This analysis indicates that the SSI technique and the inverse method, coupled to the theory of shear wave propagation in deformed hyperelastic solids, are promising and versatile tools for determining the linear and non-linear elastic parameters of brain tissue. In this analysis, we modeled the brain tissue as a non-linear *elastic* material and did not include viscosity in the data analysis. A recent study by Bercoff et al. (2004c) has demonstrated that viscosity of soft tissues can lead to the decay of the wave amplitude, but that its effect on the velocity of the shear wave is not significant provided that the attenuation length is much larger than the wavelength.

## 4. Discussions

The non-linear elastic properties of brain tissue have been measured for the first time using the SSI technique. Also for the first time, the third- and fourth-order constants *A* and *D* have been measured for soft tissue without relying on non-linear wave propagation (Rénier et al. 2008). The measurement is *easy to perform* and the data analysis is not complicated. The elastic parameters determined in this study are broadly comparable to those reported in the literature for porcine brain matter. For example, Rashid et al. (2013b) find that simple shear, destructive tests give $\mu_0$ from 1.0 kPa to 2.1 kPa and *b* from 0.1 to 0.2 for the Demiray-Fung model, while tensile tests give $\mu_0$ from 3 kPa to 5.8 kPa and *b* from 1.7 to 2.2, depending on the strain rate. Kaster et al. (2011) also use destructive testing (indentation of tissue slices) and find that $\mu_0$=0.6 kPa for white tissue and 0.4 kPa for gray tissue. Miller and Chinzei (2002) find that $\mu_0$=0.8 kPa from tensile tests. Nicolle et al. (2005) find that $\mu_0$ varies from 2.1 kPa to 16.8 kPa depending on the frequency of their oscillatory experiments. Prange and Margulies (2002) find lower values for $\mu_0$, around 200 Pa.

Our brain samples were intact but removed from the skull for *ex vivo* measurements. Between harvesting and testing, they were preserved near ice-cold conditions, to minimize the difference in mechanical response between *ex vivo* and *in vivo* testing (Rashid et al., 2013a). Nonetheless, the differences with the *in vivo* response still depend on a number of factors which were not accounted for, such as the absence of cerebral blood flow, of cerebro-spinal fluid, of pressurization, etc., see for instance the review by Chatelin et al. (2010). The characterization of the brain *in vivo* material parameters using our method could be realized by following, for example, the experimental set-up used by Miller et al. (2000) for the indentation of porcine brain tissue, or by working on trepanned animals, as performed by Macé et al. (2011).

It should be pointed out that the region of interest (circled area in Fig. 3) involved in the present measurements contains both gray matter and white matter. It remains a challenge to use the SSI technique to characterize the mechanical properties of gray and white matter separately. Some authors (Kruse et al. 2008; Green et al. 2008, Sack et al., 2009) have used the MRE method to report on the linear mechanical properties of gray and white matter. We note that *(a)* the low frequency of plane shear waves usually generated in brain tissue in these measurements gives a wavelength which is comparable or even greater than the typical size of the gray or white matter areas; and *(b)* the application of the classical relation of linear elasticity $\rho c^2 = \mu_0$ might be inappropriate for highly deformable soft matter such as brain tissue. These important issues deserve further investigation.

We note that our region of interest was a disk of diameter 4 mm, which is smaller than the sample size in previous destructive experiments conducted on brain tissue, where homogeneity of the material properties was assumed (Miller and Chinzei 1997; Donnelly and Medige 1997; Rashid et al. 2013b).

We also point out that our analysis relied on the assumption of homogeneity of the pre-strain in the region of interest. Four characteristic points, two at the upper surface and two at the bottom surfaces were used to evaluate the principal stretches along loading direction. By measuring the ratio between the final length and the initial length of a given material line, we obtain the stretches, see Jiang et al. (2014) for details. This limitation can be overcome provided that the displacements at the points around the region of interest can be accurately recorded, which is an on-going issue in ultrasonic imaging. Moreover, the lateral stretches were found to be more or less equal, so that $\xi$ could be taken as 0.5, corresponding to unconfined lateral expansion of an isotropic, incompressible solid. A deviation from that value can easily be implemented into Eqs.(9-12), but more precise measurements such as using ultrasonic speckle tracking (O'Donnell et al., 1994) are needed, especially as its value affects the deduced values of the third- and fourth-order elastic constants *A* and *D*.

Another issue that we did not address is that of intrinsic anisotropy. We refer the readers to the article by Macé et al. (2011) for pointers to further evidence of anisotropic response of the waves. In that study the brain was encased in its skull, and it is not possible to determine the origin of the anisotropy. At any rate, the combination of strain-induced and intrinsic anisotropy can be incorporated into the theory of acousto-elasticity if needed, see Destrade et al. (2010c) for instance.

## 5. Concluding Remarks

Determining the mechanical properties of brain tissue is of great importance from both scientific and technical points of view. In this paper, we make an effort to measure both the linear and non-linear elastic properties of brain tissue based on the SSI technique (Bercoff et al. 2004a) and an inverse approach developed within the theory of the propagation of small-amplitude shear waves in a deformed hyperelastic solid. Using this method, we measure the elastic and hyperelastic properties of six porcine brain tissues.

The initial shear moduli given by our measurements vary from 1.8 kPa to 3.2 kPa, the stiffening parameters *b* in the Demiray-Fung model are in the range of 0.13 to 0.73, the third-order Landau coefficients *A* are in the range -19.6 kPa to -1.3 kPa, and the fourth-order elastic constants *D* in the range 3.1 kPa to 8.7 kPa, depending on the specimen. These values are physically reasonable, in the sense that $\mu_0 > 0$, as it should be to ensure a positive initial slope in the shear response; *A<0*, as it should be to ensure strong ellipticity of the incremental equations of motion; and $\mu_0$, *A, D* are of the same order of magnitude, as expected by Destrade and Ogden (2010). These quantities may thus be used as the inputs in the simulations of the traumatic brain injury and virtual neurosurgery.

The experiments in this study are limited to the *ex vivo* measurements on an otherwise intact brain; however, *in vivo* measurements may be realized in the future with appropriate experimental set-up.

**Conflict of interest statement**

The authors have no financial and personal relationships that could inappropriately influence or bias this work.

**Acknowledgements**

Supports from the National Natural Science Foundation of China (Grant No. 11172155), Tsinghua University (2012Z02103) and 973 Program of MOST (2010CB631005) are gratefully acknowledged. We also thank the referees for helping us improve greatly previous versions of the article.

**Figure captions**

Figure 1  Elastic Cerenkov effect induced by the moving point force in a soft material with the Mach number $M_e$=3, **(a)**; and **(b)** $M_e$=30. For a large Mach number, the angle of the Mach cone is very small and the two quasi-plane shear waves form and propagate in opposite directions. The results are given by finite element analysis using an axisymmetric model.

Figure 2  A harvested porcine brain , **(a)**; and **(b)** applying the supersonic probe onto a porcine brain in water.

Figure 3  Measuring the wave speed in a porcine brain at rest (top) and compressed (bottom). The region of interest is delimited by the dotted disc (diam. around 4mm). Its plane is vertical and perpendicular to the median fissure. Its center is located approximately 9mm below the top surface.

Figure 4  Variation of the shear wave velocities with the compressive amount for the left lobes of the six different samples. The error bars show the standard deviation. The pre-strain parameter of tri-axiality $\xi$ is taken as 0.5 according to the deformation of the tested material.

Figure 5  Variation of the shear wave velocities with the compressive amount for the right lobes of the six different samples. The error bars show the standard deviation. The pre-strain parameter of tri-axiality $\xi$ is taken as 0.5 according to the deformation of the tested material.

**Figures**

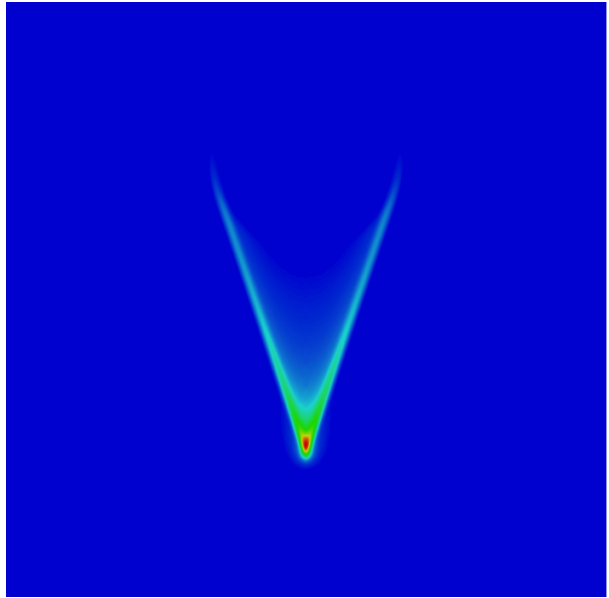

(a)

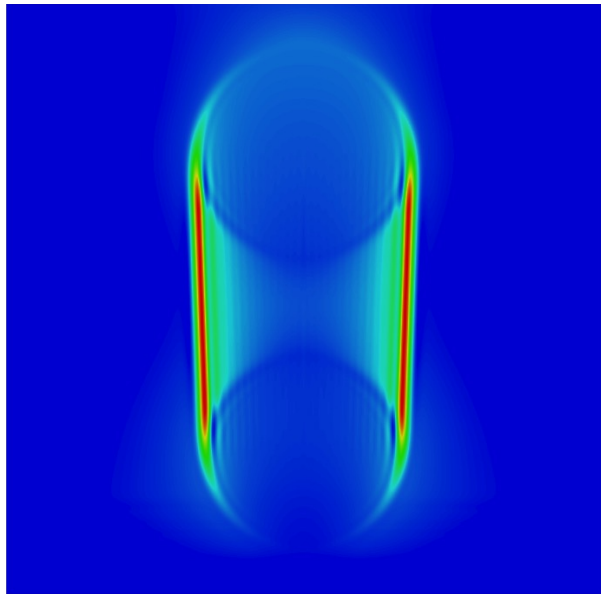

(b)

Figure 1

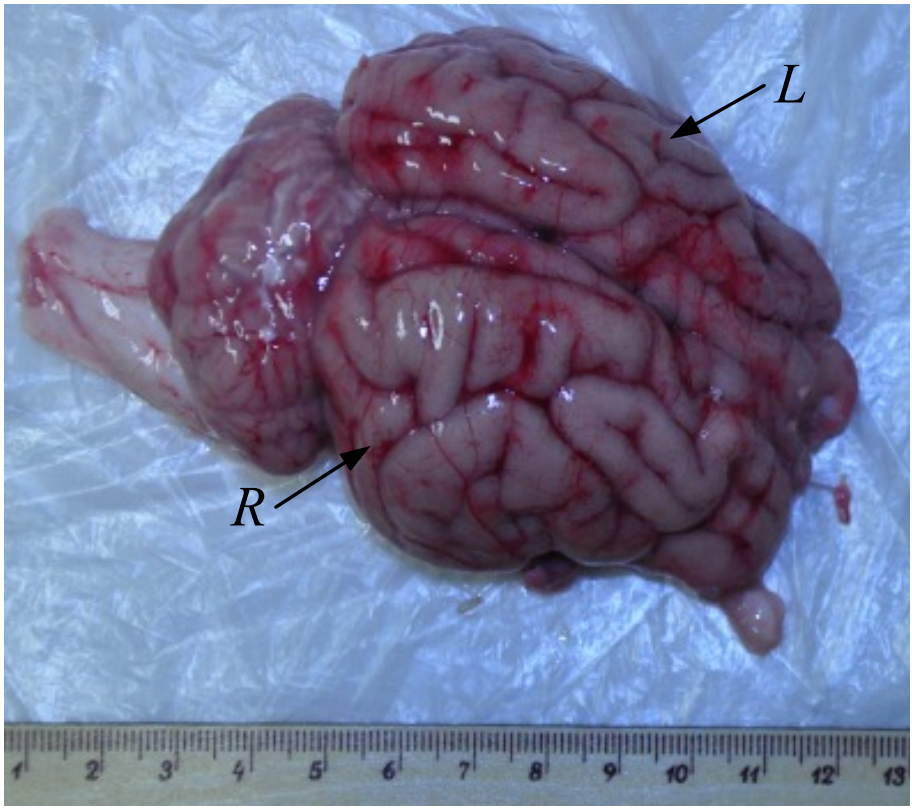

Figure 2

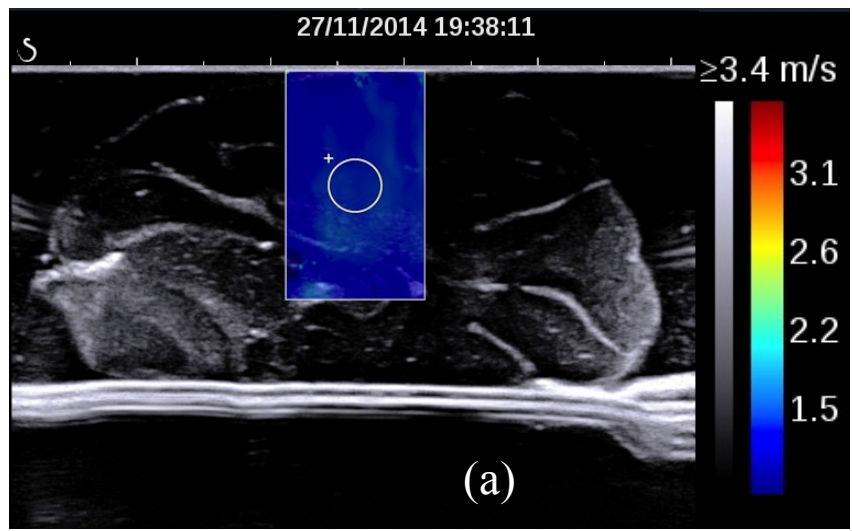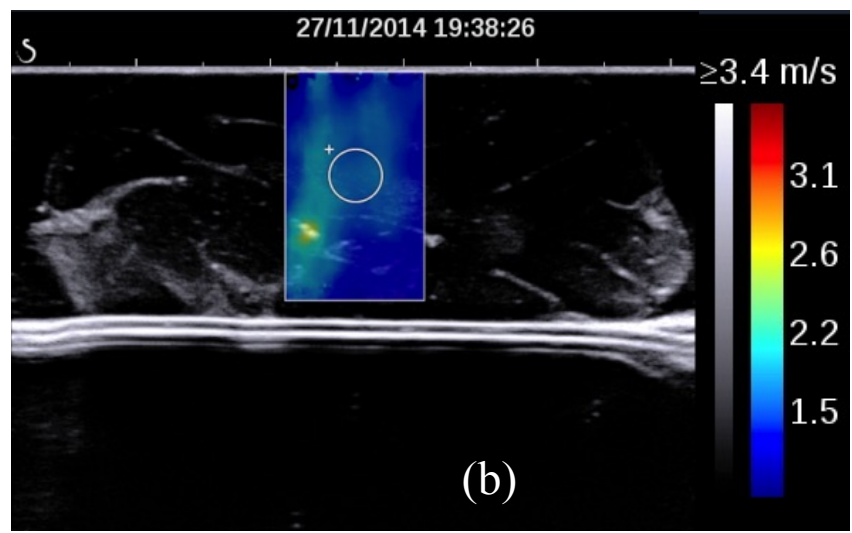

Figure 3

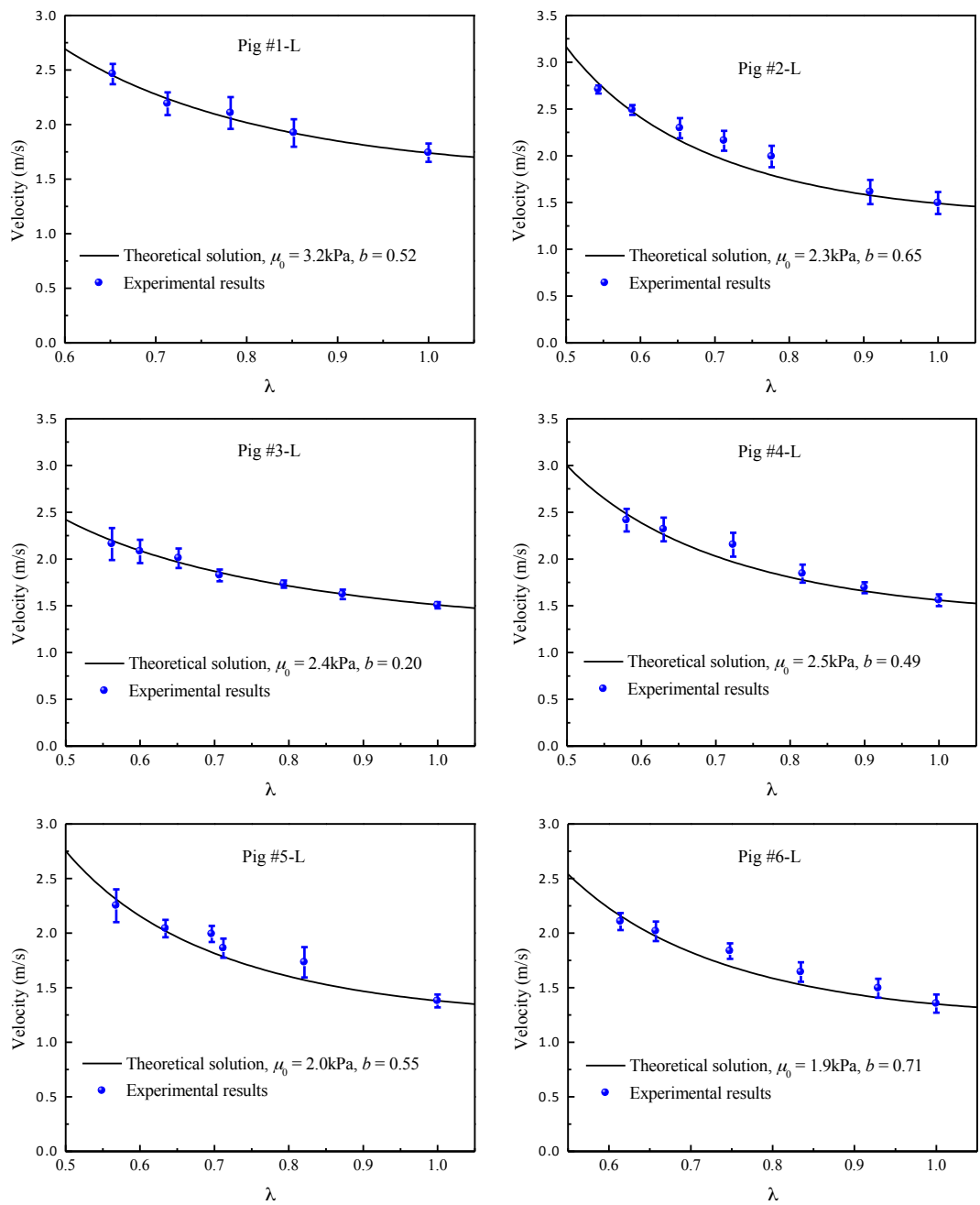

Figure 4

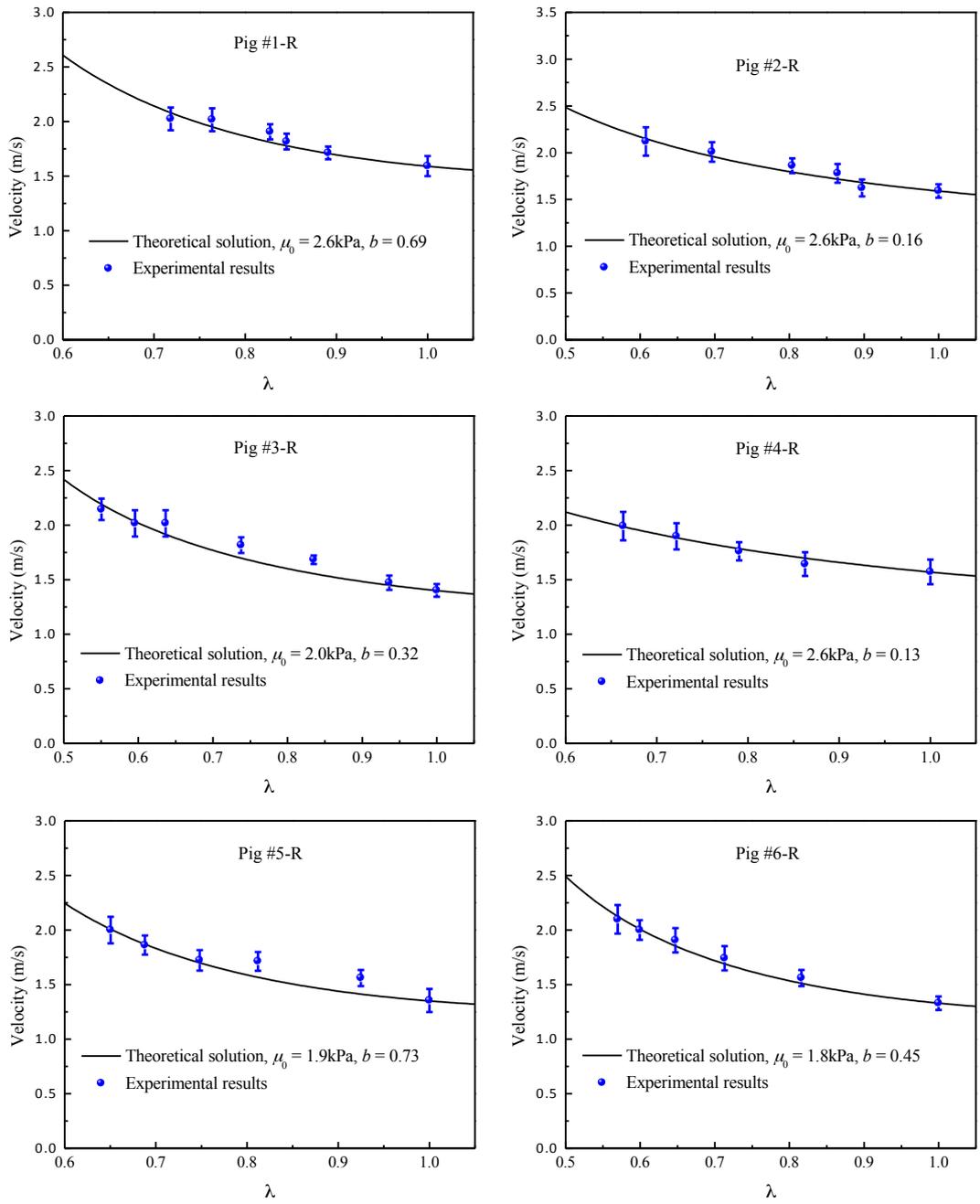

Figure 5

**Table 1**
**Initial shear modulus and non-linear elastic parameters of 6 pig brains (kPa)**

| Pig | Fourth-order elasticity | | | | | |
|---|---|---|---|---|---|---|
| | Left | | | Right | | |
| | $\mu_0$ | $A$ | $D$ | $\mu_0$ | $A$ | $D$ |
| #1 | 3.2 | -5.6 | 6.8 | 2.6 | -19.6 | 6.3 |
| #2 | 2.3 | -15.0 | 8.7 | 2.6 | -14.0 | 4.5 |
| #3 | 2.4 | -6.1 | 3.7 | 2.0 | -18.3 | 5.8 |
| #4 | 2.5 | -18.8 | 7.2 | 2.6 | -1.3 | 3.1 |
| #5 | 2.0 | -17.7 | 6.9 | 1.9 | -20.6 | 6.3 |
| #6 | 1.9 | -19.2 | 7.0 | 1.8 | -6.5 | 4.5 |

**Table 2** Average values and SD of the initial shear modulus and nonlinear elastic parameters of 6 pig brains (in kPa except for $b$, which is non-dimensional)

| | $\mu_0$ | $b$ | $A$ | $D$ |
|---|---|---|---|---|
| Left | 2.4 ± 0.5 | 0.52 ± 0.17 | −13.7 ± 6.3 | 6.7 ± 1.6 |
| Right | 2.3 ± 0.4 | 0.41 ± 0.26 | −13.4 ± 7.9 | 5.1 ± 1.3 |
| $p$ | 0.36 | 0.42 | 0.95 | 0.16 |